\documentstyle[12pt]{article}

\makeatletter
\def\eqnumsection{\@addtoreset{equation}{section}\def\theequation
{\arabic{section}.\arabic{equation}}}
\makeatother

\title{
\begin{flushright}
{\large Yaroslavl State University\\
        Preprint YARU-HE-97/01 \\
        hep-ph/9702284} \\[12mm]
\end{flushright}
       {\LARGE\bf Electromagnetic Catalysis} \\ 
       {\LARGE\bf of the Radiative Decay of the Axion}}

\author{{\Large\bf N.V.~Mikheev } \\[2mm]
        {\large\it
             Division of Theoretical Physics, Department of Physics,} \\
        {\large\it
             Yaroslavl State University, Yaroslavl 150000, Russia} \\
        {\large\it E-mail: mikheev@yars.free.net} \\[4mm]
        {\Large\bf and} \\[4mm]
        {\Large\bf L.A.~Vassilevskaya} \\[2mm]
        {\large\it
             Moscow Lomonosov University, V-952, Moscow 117234, Russia} \\
        {\large\it E-mail: vasilevs@vitep5.itep.ru}}

\date{}

\begin{document}

\maketitle

\bigskip

\begin{abstract}
The radiative decay of the axion $a \rightarrow \gamma \gamma$ is 
investigated in an external electromagnetic field in DFSZ model in
which axion couples to both quarks and leptons at tree level.
The decay probability is strongly catalyzed by the external field, 
namely, the field removes the main suppression caused by the smallness 
of the axion mass. 
\end{abstract}

\vspace*{40mm}

\centerline{\large will be published in {\it Phys. Lett. B}}

\thispagestyle{empty}

\newpage

Various extensions of the Standard Model (SM) predict the existence of 
light neutral pseudoscalars which are the subject of constant theoretical 
and experimental studies.
The axion~\cite{P1,WW} is the most widely discussed pseudoscalar
particle proposed by R.Peccei and H.Quinn to solve a major theoretical
problem of CP-conservation in strong interactions.
Although the original axion, associated with Peccei-Quinn ($PQ$) symmetry
breaking at the weak scale ($f_w$), is excluded experimentally, many 
variant $PQ$ models and their associated axions are of great interest 
(see, for example~\cite{Raf1,P2}). If the breaking scale of $PQ$ 
symmetry $f_a \gg f_w$ (the latest astrophysical data
yield $f_a \geq 10^{10} GeV$), the resulting, so called, invisible axion 
is very light ($m_a \sim f_a^{-1}$) and very weakly coupled
(coupling  $ \sim f_a^{-1}$).
At present astrophysical and cosmological considerations leave
a rather narrow window for the axion mass ~\cite{Raf1,Tur,Raf2}:

\begin{equation}
10^{-6}~{\rm eV} \leq m_a \leq 10^{-3}~{\rm eV}~.
\label{eq:MA}
\end{equation}

\noindent 
A survey of various processes involving axion production and
astrophysical methods for obtaining constraints on the parameters of 
axion models is given in~\cite{Raf1}.
Unfortunately a problem of its possible existence which is very
important for the theoretical description of elementary particles
remains unresolved. It can be naturally
explained by the fact that the axion lifetime 
in vacuum is  very large ~\cite{Raf1}:

\begin{equation}
\tau (a \to 2\gamma) \sim 6,3 \cdot 10^{48} \; s \,
\left ( {10^{-3} eV \over m_a } \right )^6 \; 
\left ( {E_a \over 1 MeV } \right ) .
\label{eq:T0}
\end{equation}

On the other hand, one can get more precise constraints on 
weakly coupled particles properties with an influence of external 
electromagnetic fields taken into account.
It is also important in some astrophysical and cosmological considerations
to take strong external fields (of order the critical, so called,
Schwinger value $F_s = m^{2}_{e}/e \simeq 0.44 \cdot 10^{14} \, G$)
into account. Notice that the decay probability in an external field 
depends on not only typical 
kinematical invariants of type $m^2$, $p^2$, but field invariants 
$\vert e^2 (F F) \vert^{1/2}$, $\vert e^2 (\tilde F F) \vert^{1/2}$,
$\vert e^2 (p F F p) \vert^{1/3}$ as well (here 
$m$ and $ p_{\mu} $  are the mass and the 4-momentum of the particles,
$ F_{\mu \nu} $ is the external field tensor, 
$ p F F p = p^\mu F_{\mu \nu} F^{\nu \rho} p_\rho$).
In ultrarelativistic limit the field invariant
$\vert e^2 (p F F p) \vert^{1/3}$ can occur the largest one.
This is due to the fact that in the relativistic
particle rest frame the field may turn out to be of order of the
critical one or even higher, appearing very close to the constant
crossed field ( $\vec E \perp \vec H, E = H$ ), where
$(FF)=(F \tilde F ) = 0$. Thus, the calculation in this field
is the relativistic limit of the calculation in an arbitrary 
relatively smooth field, possesses a great extent of generality
and acquires interest by itself.

In this work we study the double radiative axion decay 
$a \rightarrow \gamma \gamma$ in the external crossed field.

\vspace*{3mm}

We consider Dine-Fischler-Srednicki-Zhitnitskii (DFSZ)
axions~\cite{DFSZ} which couple to both quarks and leptons
at tree level.
The corresponding interaction lagrangian has the form:

\begin{equation}
{\cal L}_{af} =  - i g_{af} (\bar f \gamma_5 f) a
\label{eq:L1}
\end{equation}

\noindent where $g_{af} = C_f m_f/f_a$ is a dimensionless 
Yukawa coupling constant, $C_f$ is a model-dependent factor,
$m_f$ is a fermion mass.

A matrix element of $a \rightarrow \gamma \gamma$ decay
is described by two three-point loop diagrams (Fig.1)
where double lines imply that the influence of the external
field in the propagators of virtual fermions is taken into
account exactly. An expression for the matrix element in the
external field is:


\hspace{70mm}
\begin{figure}[tb]


\def\markphotonatomur{\begin{picture}(2,2)(0,0)
                             \put(2,1){\oval(2,2)[tl]}
                             \put(0,1){\oval(2,2)[br]}
                     \end{picture}
                    }
\def\markphotonatomdr{\begin{picture}(2,2)(0,0)
                             \put(1,0){\oval(2,2)[bl]}
                             \put(1,-2){\oval(2,2)[tr]}
                     \end{picture}
                    }
\def\photonurhalf{\begin{picture}(30,30)(0,0)
                     \multiput(0,0)(2,2){5}{\markphotonatomur}
                  \end{picture}
                 }
\def\photondrhalf{\begin{picture}(30,30)(0,0)
                     \multiput(0,0)(2,-2){5}{\markphotonatomdr}
                  \end{picture}
                 }


\unitlength=1.00mm
\special{em:linewidth 0.4pt}
\linethickness{0.4pt}

\begin{picture}(60.00,45.00)(-30.00,00.00)
\put(35.00,32.50){\oval(20.00,15.00)[]}
\put(35.00,32.50){\oval(16.00,11.00)[]}

\put(26.00,32.50){\circle*{2.00}}
\multiput(25.00,32.50)(-5.00,0.00){4}{\line(-1,0){3.00}}

\put(43.20,36.00){\circle*{2.00}}
\put(43.20,29.00){\circle*{2.00}}

\put(36.50,39.00){\line(-3,2){4.01}}
\put(36.50,39.00){\line(-3,-2){4.01}}

\put(32.50,26.00){\line(3,2){4.01}}
\put(32.50,26.00){\line(3,-2){4.01}}

\put(23.00,28.00){\makebox(0,0)[cc]{\large $x_1$}}
\put(44.00,42.00){\makebox(0,0)[cc]{\large $x_2$}}
\put(44.00,23.00){\makebox(0,0)[cc]{\large $x_3$}}

\put(15.00,35.00){\makebox(0,0)[cb]{\large $a(p)$}}

\put(35.00,45.00){\makebox(0,0)[cc]{\large $f$}}
\put(33.00,20.00){\makebox(0,0)[cc]{\large $f$}}

\put(53.00,38.00){\makebox(0,0)[lb]{\large $\gamma(q_1)$}}
\put(53.00,27.00){\makebox(0,0)[lt]{\large $\gamma(q_2)$}}

\put(43.20,36.00){\photonurhalf}
\put(43.20,29.00){\photondrhalf}

\put(70.00,32.50){\makebox(0,0)[lc]
                 {\large $+ \quad (\gamma_1 \leftrightarrow \gamma_2$)}}

\put(60.00,10.00){\makebox(0,0)[cc]{\large Figure 1.}}
\end{picture}

\end{figure}


\vspace{-5mm}

\begin{eqnarray}
{S} & = & {  e_f^2 g_{af} \over \sqrt {2 E_a V \cdot 2 \omega_1 V 
\cdot 2 \omega_2 V}}
\int d^4 x_1 \, d^4 x_2 \, d^4 x_3 
\nonumber \\ 
& \times & exp \lbrace -ipx_1 + iq_1x_2 + iq_2x_3 \rbrace \\
& \times & Sp \lbrace S^{(F)}(x_1,x_3) (\varepsilon_2 \gamma )
S^{(F)}(x_3,x_2) (\varepsilon_1 \gamma ) S^{(F)}(x_2,x_1) \gamma_5 
\rbrace 
\nonumber \\ 
& + &  ( \varepsilon_1, q_1 \leftrightarrow \varepsilon_2, q_2 )
\nonumber  
\label{eq:S1}
\end{eqnarray}

\noindent Here $e_f = e Q_f$, $e > 0$  is the elementary charge,
$Q_f$ is a relative fermion charge in the loop;
$q_{1}$, $q_{2}$ and $\varepsilon_1$, $\varepsilon_2$ are 
the 4-momenta of the final photons and their polarization 4-vectors,
respectively; $p$ is the 4-vector of the decaying axion;
$(\varepsilon \gamma ) = (\varepsilon^\mu \gamma_\mu )$,  
$\gamma_\mu$, $\gamma_5$
are  Dirac $\gamma$-matrices.

\noindent The expression for the propagator of
a charged fermion in the crossed field $ S^{(F)} (x, y)$
in the proper time formalism has the form:

\begin{eqnarray}
\label{eq:SFA}
S^{(F)} (x,y) & = & e ^{\mbox{\normalsize $i \Omega (x,y)$}} S (X), \\
\Omega(x,y) & = & - e_f \, \int_y^x d\xi_\mu \, \left [ A_\mu (\xi) +
            \frac{1}{2} F_{\mu \nu} (\xi - y)_\nu \right ] ,
\nonumber \\
S (X) & = & - {i \over 16 \pi^2} \int\limits_0^\infty
{ds \over s^2} \bigg [ {1 \over 2s} (X \gamma) -
{i e_f \over 2} (X \tilde F \gamma) \gamma_5
\nonumber \\
& - & {s e_f^2 \over 3} (X F F \gamma) + m_f
+ {s m_f e_f \over 2} (\gamma F \gamma) \bigg ]
\nonumber \\
& \times & \exp \left [ -i \left ( s m_f^2 + {X^2 \over 4 s} +
{s e_f^2 \over 12} (XFFX) \right ) \right ] , 
\nonumber
\end{eqnarray}

\noindent where $X = x - y$, $ m_{f}$ and $e_f$ are the mass and
the charge of the intermediate fermion; $F_{\mu \nu}$ and
${\tilde F_{\mu \nu}} = (1/2) \varepsilon_{\mu \nu \alpha \beta} 
F_{\alpha \beta}$ are the tensor and the dual tensor of the constant 
crossed field; $A_\mu$ is the four-potential.

Calculations of some processes amplitudes describing by
three-point loop diagrams are very complicated
\footnote{The history of calculations of an analogous 
three-point loop process of the photon splitting
$\gamma \to \gamma \gamma$ in external electromagnetic fields
goes back to the pioneer work by Adler ~\cite{Adl1} and it is still 
in progress ~\cite{Adl2}.} and the general expression 
for the $a \gamma \gamma$-vertex has a 
cumbersome form and will be published elsewhere. 

The field-induced contribution $\Delta M = M - M_0 $ to
the amplitude $M$ ($M_0$  is the vacuum amplitude) 
of the ultrarelativistic axion decay ($E_a \gg m_a$)
we have obtained is:

\begin{equation}
\Delta M  \cong   \frac{\alpha}\pi \sum_f  \frac{Q^2_f g_{af}}{m_f^5}
 \left [ \frac{e^2_f}{1 - \lambda} (f_1 \tilde F) (f_2 F) J
 +  O\left(m_a\over E_a \right)
 +  ( \varepsilon_1, q_1 \leftrightarrow \varepsilon_2, q_2 )
\right ] ,
\label{eq:M2}
\end{equation}

\begin{eqnarray}
J & = & -\int\limits_0^1 dx \int \limits_0^{1-x} dy
y (1 - y -2x) \eta^3 {d^2 f(\eta) \over d \eta^2}, 
\nonumber \\
f(\eta) & = & i \int\limits_0^\infty dt \exp \left [ - i 
\left ( \eta t + {1\over 3} t^3 \right ) \right ],  
\nonumber \\
\eta & = & (\chi^2 \; Z (x,y,\lambda))^{-1/3}, 
\nonumber \\
Z (x,y,\lambda) & = & - 2 xy( 1 - 3x - 3y + 2x^2 + 2y^2 + 3xy ) \;
\lambda (1 - \lambda)
\nonumber \\
 & + &  x^2 ( 1 - x)^2 \; (1 - \lambda)^2
+  y^2 ( 1 - y)^2 \; \lambda^2
\nonumber \\
\chi^{2} & = & \frac{e_f^2 (p  F F p)}{m_f^6}, 
\nonumber \\
f_{i \alpha \beta} & = & q_{i\alpha} \varepsilon_{i\beta} - 
q_{i\beta} \varepsilon_{i\alpha} ,
\qquad i = 1, 2,
\nonumber
\end{eqnarray}

\noindent where  $\lambda = \omega_1 / E_a$ is the first photon
relative energy, $E_a$ is the decaying axion energy; $J$ is the 
integral of the Hardy-Stokes function $f(\eta)$, $\chi$ is the,
so called, dynamic parameter. 

Using the standard methods we get an expression for the 
axion decay probability in the external field:

\begin{eqnarray}
W^{(F)}  =  {1 \over \pi E_a}\left( {\alpha \over \pi}\right )^2
\int \limits_{0}^1 d \lambda \;\lambda^2 \;\left \vert \sum \limits_f
Q^2_f g_{af} m_f \chi^2 \; J \right \vert ^2,
\label{eq:WF}
\end{eqnarray}

\noindent With the strong hierarchy of fermion masses
the contribution of the fermion with
the maximum value of the dynamic parameter 
$\chi^{2}$ can dominate in ~(\ref{eq:WF}):

\begin{eqnarray}
W^{(F)} \cong  3.32 \left({\alpha \over \pi}\right)^2 
{(Q_f^2 g_{af} m_f)^2 \over \pi E_a} P (\chi).
\label{eq:RP}
\end{eqnarray}

\noindent The numerical values of the function  $P(\chi)$ are 
presented in Table and asymptotic behaviors for small and 
large values of the dynamical parameter $\chi$ are given:

\begin{eqnarray}
{P(\chi) \bigg \vert _{\chi \ll 1} } & \simeq & 2.31 \cdot
10^{-5} \cdot \chi^8 ( 1 + 9.5\chi^2 + \cdots ),
\nonumber \\
{P(\chi) \bigg \vert _{\chi^{1/3} \gg 1} } & \simeq &
1 - {8.8 \over \chi^{1/3}} + {32.0 \over \chi^{2/3}}
- \cdots
\nonumber
\end{eqnarray}

Let us compare the probability (\ref{eq:RP}) we have obtained 
with the well known axion decay probability in vacuum~\cite{Raf1}:

\begin{eqnarray}
W^{(0)} = \frac {g_{a\gamma}^2 m_a^4 }{64 \pi E_a}.
\label{eq:A0}
\end{eqnarray}

\noindent Here $g_{a \gamma} = (\alpha / 2 \pi f_a ) 
(E/N - 1.92 \pm 0.08)$~\cite{Raf1}, 
where $E$ and $N$ are the model-dependent coefficients
of the electromagnetic and color anomalies. The comparison 

\begin{eqnarray}
R_f = {W^{(F)} \over W^{(0)}} \cong  2,12 \cdot 10^2 \; 
\left ( {\alpha \over \pi} \right )^2 \;
\left ( {Q^2_f g_{af} m_f \over g_{a \gamma} m^2_a} \right )^2 \; 
P(\chi)
\label{eq:R1}
\end{eqnarray}

\noindent illustrates the enhancing (catalyzing) influence of 
the external field on the ultrarelativistic axion decay 
($E_a \gg m_a$), because $W^{(F)}$ ~(\ref{eq:RP}) does not contain 
a suppression associated with the smallness of the decaying axion mass. 
Notice that the analogous phenomenon of strong catalyzing effect
of external electromagnetic field on rare loop process of
neutrino radiative decay was discovered in ~\cite{L1}. 
For the electron $R_e$ has the form:

\begin{eqnarray}
R_e \simeq 10^{37} 
\left ( {\cos^2\beta \over E/N - 1.92} \right )^2 \;
\left ( {10^{-3} eV \over m_a } \right )^4 \; P(\chi).
\label{eq:R2}
\end{eqnarray}

\noindent Here $\cos^2 \beta$ determines the electron Yukawa coupling 
$g_{ae} = \frac {1}{3}\cos^2\beta\; (m_e/f_a)$ in the DFSZ model.
Large values of the dynamic parameter $\chi \gg 1$, when 
$P(\chi) \sim 1$, can be realized, for example, in case of the decaying
axion energy $\sim 10$ MeV and the magnetic field strength
$\sim F_s$. At present a possible existence of strong external 
fields $F \ge F_s$ is widely discussed in astrophysics,
where strong magnetic fields can take place (a process of a 
coalescence of neutron stars, an explosion of a supernova 
of the type SN 1987A).

Note that the expressions for the 
amplitude~(\ref{eq:M2}) and the decay probability~(\ref{eq:RP})
can be applied to other pseudoscalar particles with the coupling
of type~(\ref{eq:L1}), even though these particles are massless.
This is due to the corresponding dispersion low in the magnetic
field.

\vspace*{5mm}

\noindent 
{\bf Acknowledgements}  

\vspace*{3mm}

The authors thank V.A.~Rubakov for fruitful discussions and 
M.E.~Shaposhnikov and \newline
A.Ya.~Parkhomenko for useful critical remarks.

\begin{table}[htb]
\caption{}
\begin{center}
\begin{tabular}{|c|c|}\hline
& \\
$\chi$  &  $P(\chi)$ \\
& \\ \hline
& \\
$ \qquad 10^{-2}\qquad $ & $\qquad 2.31 \cdot 10^{-21} \qquad$ \\
& \\
$10^{-1}$ & $2.53 \cdot 10^{-13}$ \\
& \\
$1$ & $2.12 \cdot 10^{-5}$ \\
& \\
$5$ & $4.68 \cdot 10^{-3}$ \\
& \\
$10$ & $1.7 \cdot 10^{-2}$ \\
& \\
$100$ & $0.17$ \\
& \\
$10^3$ & $0.44$ \\
& \\
$10^4$ & $0.66$ \\
& \\
$10^6$ & $0.871$ \\
& \\
$\infty$ & $1$ \\
& \\ \hline
\end{tabular}
\end{center}
\end{table}

\end{document}